# Transkingdom Networks: A Systems Biology Approach to Identify Causal Members of Host-Microbiota Interactions

**Authors and affiliations:**

**Richard R. Rodrigues**

College of Pharmacy, Oregon State University, 1601 SW Jefferson Way, Corvallis, Oregon 97331, USA

**Natalia Shulzhenko**

College of Veterinary Medicine, Oregon State University, 105 Dryden Hall, 450 SW 30th Street, Corvallis, Oregon 97331, USA

**Andrey Morgun**

College of Pharmacy, Oregon State University, 1601 SW Jefferson Way, Corvallis, Oregon 97331, USA

**Corresponding authors:**

Andrey Morgun andriy.morgun@oregonstate.edu

Richard R. Rodrigues rodrrich@oregonstate.edu

**Running head:**

Transkingdom Network Analysis

# Transkingdom Networks: A Systems Biology Approach to Identify Causal Members of Host-Microbiota Interactions

Richard R. Rodrigues, Natalia Shulzhenko, Andrey Morgun

**Abstract**

Improvements in sequencing technologies and reduced experimental costs have resulted in a vast number of studies generating high-throughput data. Although the number of methods to analyze these "omics" data has also increased, computational complexity and lack of documentation hinder researchers from analyzing their high-throughput data to its true potential. In this chapter we detail our data-driven, transkingdom network (TransNet) analysis protocol to integrate and interrogate multi-omics data. This systems biology approach has allowed us to successfully identify important causal relationships between different taxonomic kingdoms (e.g. mammals and microbes) using diverse types of data.

**Introduction**

Over the last decade assessing eukaryotic and prokaryotic genomes and transcriptomes have become extremely easy. With technologies like microarrays and next-generation sequencing, investigators now have faster and cheaper access to high-throughput "-omics" data [1]. This in turn has increased the number of analysis methods [2] and allows for the exploration of new and different biological questions to provide insights and better understanding of host, host-microbial systems, and diseases [3-5].

Studies usually focus on identifying differences between "groups" (e.g., healthy versus diseased or treatment versus control) or changes across a time course (e.g., development of an organism or progression of a disease). Depending on the biological questions, such studies generate one or more types of omics data [6-8], e.g., host gene expression and gut microbial abundance. Typically, studies analyze these omics data separately, comparing gene expression and microbial abundance between groups or across stages. Although such analysis methods have been very useful, they do not directly answer the most critical questions of host-microbiota interactions, i.e. which microbes affect specific pathways in the host and which host pathways/genes control specific members of the microbial community? Therefore, to answer those questions, these analyses are usually followed by literature searches to identify relationships between host genes and microbes.

Different algorithms and methods have been proposed to integrate multi-omics data [9-13]. More recently, a few published studies have not only integrated microbiome and host data, but have also been able to successfully test their computational predictions in the laboratory [14-19]. In this chapter we describe our data-driven, transkingdom network (TransNet) analysis pipeline (Figure 1) that has allowed us to make validatable computational inferences. We construct networks using correlations between differentially expressed elements (e.g., genes, microbes) and integration of high throughput data from different taxonomic kingdoms (e.g., human and bacteria). In fact, TransNet analysis can be applied to integrate any "Transomics" data, between as well as within taxonomic kingdoms e.g., miRNA and gene expression, protein and metabolite, bacterial and host gene expression, or copy number, methylation, and gene expression. Interrogation of this network allows us to pinpoint important causal relationships between data. For example, using this method we inferred and validated: 1) microbes and microbial genes controlling a specific mammalian pathway [15]; 2) a microbe that mediates effect of one host pathway on another [14]; 3) a host gene that mediates control of gut microbe through an upstream master regulator gene [14]. Below we show how TransNet analysis can be used to integrate host gene expression with microbial abundance to create transkingdom networks.

**Materials**

**Program Availability:**

Our transkingdom network analysis pipeline is independent of programming language or software. However, for ease of access and usage simplicity, we have provided our pipeline as a convenient R package ([TransNetDemo](#)) and supplementary document (File S1) in addition to the description provided. Although the user can choose to perform the following steps in a programming language or software of their choice, we suggest using our R package.

*Required R packages:*

Install the following packages along with their dependencies: stringr, ProNet, igraph, ggplot2, gplots from CRAN (https://cran.r-project.org/).

*Installing TransNetDemo:*

library(devtools)

install_github("richrr/TransNetDemo")

library(TransNetDemo)

**Data Sources:**

Due to a variety of data generation technologies, biological questions, and software, description of every possible analysis is beyond the scope of this chapter. We expect that the user has access to tab-delimited file(s) containing the measurements of biological data type, e.g., gene expression, copy number, methylation, miRNA, or microbial across samples. Depending on the data type the user can find reviews and protocol papers describing the analysis needed to produce "abundance" tables [20-24].

The transkingdom network analysis method can be applied to any experimental design (e.g. treated/untreated, control/disease). As an example we will use simulated data from a simple experimental design that investigates the effects of factors on host-microbial interactions. Consider an experiment with two groups (HFHS/NCD) of 25 samples each. At the end of the experiment, among other phenotypic measurements (e.g., body weight, enzyme levels, hormone levels), the gene expression levels and microbial abundance in the gut (e.g., ileum) of the samples were measured. Depending on resource availability, high confidence and consistent results can be achieved by increasing the number of samples per group and/or repeating the above experiment multiple times. In this example data, we have two such experiments. A brief description of how to generate the abundance tables is mentioned below. Information about how the network analysis protocol can be adapted to answer some other biological questions have been mentioned in the NOTES section of this chapter.

**Gene expression analysis:**

Several different technologies, each with their own pros and cons, allow for the measuring of transcriptome levels in an organism. Although microarrays were extensively used over the last two decades, the availability of cheap and efficient library preparation kits and sequencing methods allow for the expression measurements of known and novel genes using RNA-Seq technologies [25].

In case of RNA-Seq data, the sequencing facilities usually provide fastq files that contain raw reads per sample (demultiplexed). Here, the number of reads corresponding to a particular gene is proportional to that gene's expression level. Software like FASTX-Toolkit (http://hannonlab.cshl.edu/fastx_toolkit/), FastQC (http://www.bioinformatics.babraham.ac.uk/projects/fastqc/), PRINSEQ [26], or cutadapt [27] can be used for adapter removal and quality control. Depending on the availability of a gold-standard reference host genome sequence, gene expression abundance can be measured using the Tuxedo [21] or Trinity [28] pipeline. Both of these pipelines permit the analysis of single or paired reads and different read lengths while outputting a file containing the expression levels (number of reads) of genes (rows) present in each sample (columns). The obtained read counts can be normalized by simple (e.g., quantile normalization, total reads (CPKM), reads per kilobase per million mapped reads (RPKM) [29]) or sophisticated methods (e.g., DESEQ [30], edgeR [31]).

In the case of microarray data, hybridization facilities usually provide scan files (Affymetrix CEL, Illumina IDAT, or GenePix GPR) that contain the intensity of probes per sample. Here, the probe intensity is proportional to the corresponding gene expression level. Software like Affymetrix® Expression Console™, Illumina's GenomeStudio, and GenePix® Pro, as well as packages like affy [32] and limma [33], allow for background correction, normalization, and summarized probe intensities while outputting a file containing the expression levels of genes (rows) present in each sample (columns).

**Microbial abundance analysis:**

The abundance of next-generation technologies has helped in the study of microbial richness and diversity. Scientists no longer need to rely on cultivation methods and can directly sequence the microbiome, helping explore previously unknown microbes. The amplicon based sequencing technologies rely on using a gene marker (16S [34,35] ribosomal RNA gene, Internal Transcribed Spacer [36,37], etc.) to identify microbial presence and abundance. Although relatively cheaper than the shotgun metagenomics, they rely on databases of known genomic markers to identify microbes and rarely provide taxonomy at the species or strain levels. The shotgun metagenomics sequencing approach does a better job at surveying the entire genome of microbes since it does not focus on amplifying specific genes. Consequently, it provides fine-grained taxonomic information along with a more accurate representation of the microbial structure and function, including the previously unknown "dark matter" microbes [38].

Software like QIIME [39], MOTHUR [40], etc. provide all-in-one toolkits that can demultiplex, perform quality control, and analyze the amplicon based sequences. Similar to RNA-Seq data, the fastq files obtained from the sequencing facility need to be processed for the removal of barcodes, adapter, and primers followed by filtering to retain high quality sequences. The reads are grouped (binned) per sequence similarity (usually at 97% threshold) into operational taxonomic units (OTUs). The taxonomy of a known microbe (or the ancestor taxonomy of the top matches) closest to the representative sequence of the OTU is assigned to all the reads in that OTU. The tools output a file containing the abundance

(number of reads) of OTUs (rows) present in each sample (columns). The obtained read counts can be relativized or cumulative sum scaling (CSS) [41] normalized.

Shotgun metagenomic data can be analyzed [38] using tools such as MG-RAST [42], MEGAN [43], MetaPhlAn [44], and HUMAnN [45]. Although most of these software provide taxonomic and functional analyses, they are not standalone. Demultiplexing and quality control need to be done before the reads are imported in the software. Especially in case of host-microbe systems, Processing utility for Metagenomics Analysis (PuMA) (http://blogs.oregonstate.edu/morgunshulzhenkolabs/softwares/puma/) provides an all-inclusive software pipeline that can be more user-friendly. PuMA uses cutadapt for quality control and Bowtie [46] to identify reads that match the host genome and discards these "contaminating" reads from downstream analysis. The remaining microbial reads are aligned to a database of known protein sequences using DIAMOND [47], followed by taxonomic and functional (e.g. SEED, COG, KEGG) assignments using MEGAN. PuMA outputs a file containing the abundance of microbes and pathways (rows) in each sample (columns). The appropriate normalization techniques from the RNA-Seq or amplicon sequencing methods can be performed on the abundance table.

In summary, the user needs at least one of each of the following files before starting network analysis:

- mapping file: tab-delimited file containing the group (e.g. treated/untreated, control/disease) affiliation for each sample with "Factor" and "SampleID" as column headers, respectively.
- data files: tab-delimited files containing the abundance of elements (host genes and microbes) per sample, where the elements and samples are rows and columns, respectively. Importantly, each sample must have both types of data available.
    - normalized gene expression file: the column "IdSymbol" contains the unique genes while the remaining columns contain their expression levels across different samples.
    - normalized otu abundance file: the column "IdSymbol" contains the unique microbes while the remaining columns contain their abundance across different samples.

**Methods**

The following steps will help to identify key elements of a system from high confidence modules of a multi-omics network. We show the first few steps with the gene abundance file(s) using the code from the GeneDemo.R (GD) file available in our package. It is straightforward to run similar steps on the microbe abundance file(s), however we have also provided the code in MicrobeDemo.R (MD) file for ease of use.

- Start by setting defaults for variables that you will use in the analysis, such as significance thresholds (GD: lines 7-9), groups to be compared (GD: lines 11-13), and headers of relevant columns from the mapping (GD: lines 14-15) and abundance files (GD: line 16).
- Next you want to identify the differentially expressed elements (GD: line 29). The network analysis can be performed using all the elements (genes, microbes, etc.). However, we suggest identifying the elements that show differential abundance between groups, using code from Compare_groups.R (Cg) file, to focus on the most important elements and make the analyses computationally efficient.
    - Read from the mapping file to extract the samples from each group (Cg: lines 11-20).
    - Read from the gene abundance file (Cg: lines 22-25).
    - Then perform test for differential abundance using code from Diff_abundance.R (Da) file. This function returns the mean, median for each group along with the fold change and p-value (Da: lines 8-28).
    - Next, account for multiple testing using Benjamini-Hochberg's FDR calculation (Cg: lines 38-41).
    - Finally, select the differentially expressed genes using appropriate FDR cutoff (GD: line 33) (< 0.05) (Figure 2b).

- We highly recommend that if you have datasets obtained from replicate experiments or in different sample cohorts that you perform the above steps for each experiment and do meta-analysis [16,18,48,49] (GD: lines 39-47).
    - First, the meta-analysis selects for genes that show fold change direction consistency across datasets (Check_consistency.R), e.g., upregulated (or downregulated) across all experiments.
    - Second, for the genes showing consistent fold change direction use Fisher's method to calculate a combined p-value (Calc_combined.R) from the individual p-values (from comparison test) across multiple experiments.
    - Then apply appropriate significance thresholds (Apply_sign_cutoffs.R) based on individual p-value (< 0.3) in each dataset, combined (Fisher's) p-value across datasets (< 0.05), and FDR (< 0.1) across the combined p-values to identify consistently differentially abundant elements.
    - Ensuring the same direction of regulation in all datasets and restricting individual p values at each individual dataset allows controlling of heterogeneity between datasets. Note that mere calculation of Fisher p-value for meta-analysis followed by application of FDR is not sufficient for accurate identification of differentially abundance/expression.
- Determining associations between elements (e.g., genes and/or microbes) is central for network reconstruction. Defining strength and sign of correlation (GD: line 56) can help to determine whether two elements (i.e. biological entities represented by nodes in a network) have a positive or negative interaction. Such information about potential relationships, using code from Correlation_in_group.R (Cig) file, is crucial for interrogating and understanding the regulatory mechanisms between elements. Note, correlations are calculated using data from samples representing one group (phenotypic class), never pooling samples from all groups for estimation of correlation. Therefore, the following steps should be performed for each group separately.
    - Read from the mapping file to extract the samples from a group (Cig: lines 11-18).

- 
    - Read from the gene abundance file (Cig: lines 20-23).
    - Then create pairs (Cig: lines 25-32) from the consistent genes obtained in the previous step.
    - Next perform test for correlation on gene pairs using code from Calc_cor.R (Ccr) file. This function returns the correlation and p-value (Ccr: lines 8-17).
    - Next, account for multiple testing using Benjamini-Hochberg's FDR calculation (Cig: lines 48-50).
    - Finally, select the significantly correlated gene pairs using appropriate FDR cutoff (GD: line 60) < 0.1.
- We highly recommend that if you have datasets obtained from replicate experiments or different sample cohorts that you perform the above steps for each experiment and do meta-analysis (GD: lines 65-72).
    - First, the meta-analysis selects for gene pairs that show correlation direction consistency across datasets (Check_consistency.R), e.g., positive (or negative) across all experiments.
    - The next steps of combining the individual p-values (from correlation test) and applying multiple significance cutoffs are similar to those in the meta-analysis of genes.
- At this point you have a network for a single group where nodes are genes and edges indicate significant correlation. Next, we identify the proportion of unexpected correlations (PUC) [50] (GD: line 83). If two elements have a regulatory relationship we expect them to behave in certain ways. For example, consider two groups. Two positively correlated genes in a group should have the same direction of fold change between two groups. On the other hand, two negatively correlated genes should have the opposite direction of fold change. Edges in a network where the sign of correlations does not correspond to the direction of change are unexpected, are not likely to contribute to the process under investigation, and hence, discarded using code from Puc_compatiable_network.R (Pcn) file.
    - First, for each gene pair identify the sign of correlation (Pcn: lines 47-53).

- - Second, calculate if each gene in the pair has the same direction of regulation (i.e. fold change) (Pcn: lines 56-65).
  - Pairs are expected and kept (Pcn: lines 70-80) if they satisfy either of these conditions:
    - positively correlated genes have the same fold change direction
    - negatively correlated genes have different fold change direction
- At this point you have a network that satisfies regulatory relationships. Next, the obtained network can be systematically studied to answer different biological question. Most often, network interrogation relies on identifying highly inter-connected sets of nodes. Such a sub-network is called a module (or cluster). Identify clusters (GD: line 89) using the MCODE method from the Identify_subnetworks.R file (Figure 3b).
- Repeat the above steps for the microbial (or any other data type) abundance file(s) to obtain heat map (Figure 2a) and clusters (Figure 3a) per biological data type (e.g. genes, microbes, etc.). Refer to the code in MicrobeDemo.R file.
- The next step is to integrate sub-networks to create transkingdom networks using code from the GeneMicrobeDemo.R (GMD) file. Note that at this point you have already identified modules from the gene and the microbe networks. Similar to the above steps, create pairs between nodes from the different modules (GMD: line 29), calculate correlations within a group (GMD: line 32), and identify significant pairs based on single (GMD: line 36) or meta (GMD: lines 41-49) analysis. Next, apply PUC analysis and remove unexpected edges from this transkingdom (gene-microbe) network as it is done for regular gene expression (and microbial abundance) network (GMD: line58).
- Combining the gene-gene correlations (edges from the gene sub-networks) (GMD: line 74), microbe-microbe correlations (edges from the microbe sub-networks) (GMD: line 77), and the gene-microbe correlations (GMD: line 80) creates the full transkingdom network (GMD: line 83) (Figure 4).

- Finally, identify elements that are crucial for crosstalk between the different modules in a network using bipartite betweenness centrality (BBC) (GMD: lines 92-119). This approach involves calculating the shortest paths between nodes from different modules using code from Get_shortest_paths.R file. The elements with the highest BBC measurement (GMD: lines 123-128) are more likely to be critical in mediating the transfer of signals between the different modules of a network and candidates for further experimentation.

**Notes**

The above protocol was written for a step-by-step introduction to transkingdom network analysis. Although the above experimental setup and analyses should suffice in most cases please see the following suggestions for other alternatives to the analysis.

- Data normalization is a crucial step in analysis and network reconstruction [51], hence choose the appropriate normalization method for your biological data [52,53]. No normalization method universally out-performs other methods. However, if unsure about which normalization to use we recommend quantile normalization followed by log transformation since, in our experience, it works well for most biological data.
- Depending on the experimental design and biological question apply appropriate parametric (paired or unpaired t-test, analysis of variance (ANOVA), multivariate ANOVA (MANOVA), etc.) and non-parametric (Man-Whitney, Wilcoxon rank sum test, Multi-response Permutation Procedures (MRPP), etc.) tests to identify differential abundance.
- It is common practice to visualize the levels of differentially abundant elements. The code from Heatmaps.R file can help to visualize the significant genes and microbes from our example.

- Depending on sample size a Pearson or Spearman correlation between two elements from the same samples should suffice. However, use partial correlation [54] or other methods [55] to detect correlations and reduce indirect interactions.
- The network analysis can be extended to identify differentially correlated genes in co-expression networks obtained for the different groups and uncover regulatory mechanisms in phenotypic transitions [56,57].
- Cfinder and graph clustering (MCL) [19] are other tools to help identify modules in networks.
- We can also inspect multiple network topology properties such as the degree and centrality measures using *NetworkAnalyzer* in Cytoscape to identify important elements in the full transkingdom network.

**Acknowledgements**

The authors thank Karen N. D'Souza, Khiem Lam, and Dr. Xiaoxi Dong for their help in writing the book chapter. This work was supported by the NIH U01 AI109695 (AM) and R01 DK103761 (NS).

**Figure legends:**

**Figure 1: Overview of transkingdom network analysis.** Omics data for multiple data types (e.g. microbial, gene expression, etc.) are analyzed to identify differentially abundant elements (e.g. microbes, genes, etc.). For each group (e.g. treatment or control) co-expression networks are constructed for each data type followed by the identification of dense sub-networks (modules). Calculating correlations between module elements of the different data types creates the "transkingdom" network. Network interrogation of the transkingdom network allows identification of causal members and regulatory relationships.

**Figure 2: Heat map from hierarchical clustering of differentially abundant elements.** Rows indicate (a) microbes and (b) genes, while columns indicate samples. The pink and blue colors indicate samples belonging to the groups A (HFHS) and B (NCD), respectively. The green and red colors indicate increase and decrease, respectively, in expression or abundance, whereas brightness indicates higher fold change.

**Figure 3: Clusters obtained from the correlation networks.** The PUC compatible (a) microbe and (b) gene networks for an individual group (HFHS) are mined to identify densely connected sub-networks. Edges indicate significant correlation between elements.

**Figure 4: Transkingdom network**. A full network, for the HFHS group, contains gene-gene, microbe-microbe, and gene-microbe edges. Edges indicate significant correlation between elements. The blue and pink colors indicate gene and microbe nodes, respectively. The labeled node has the highest BBC measurement among microbes and is therefore considered to be important and a potential causal player in the experiment.

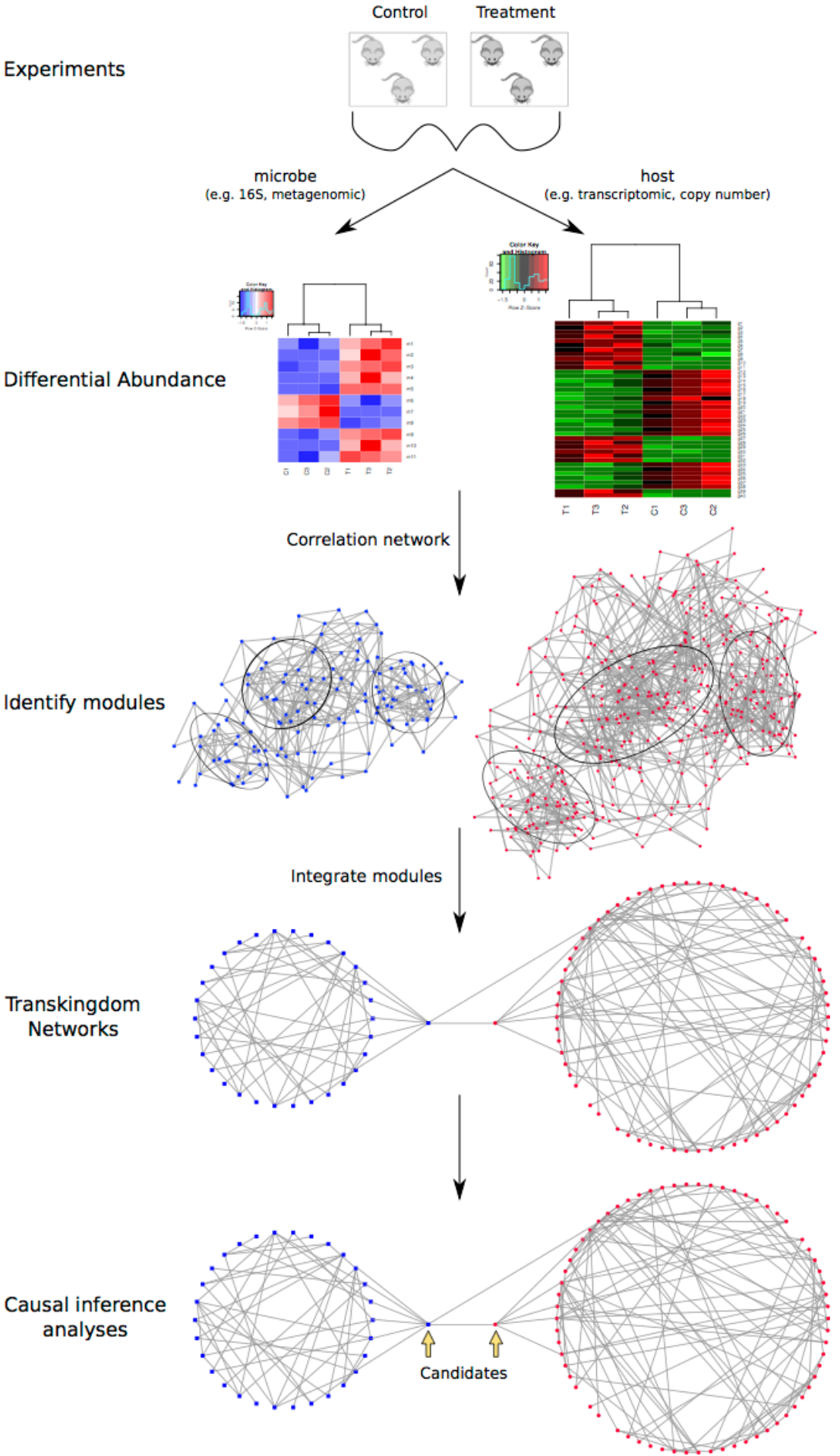

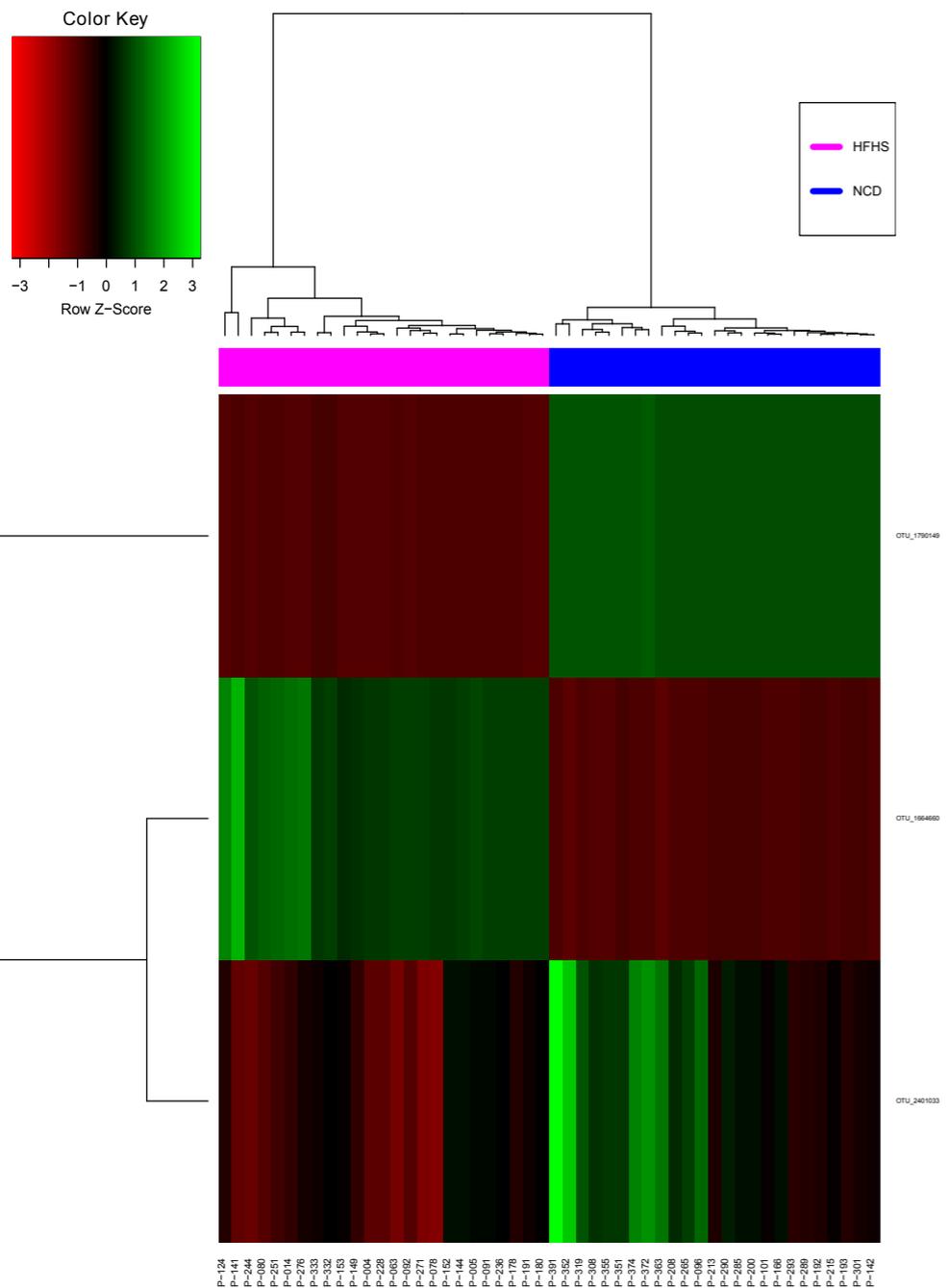
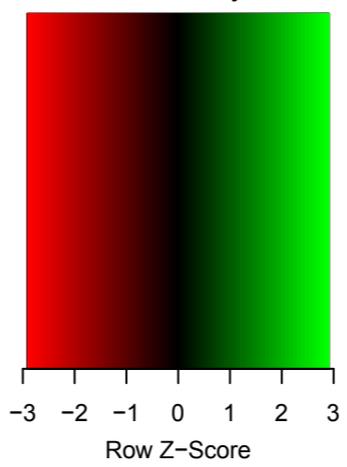
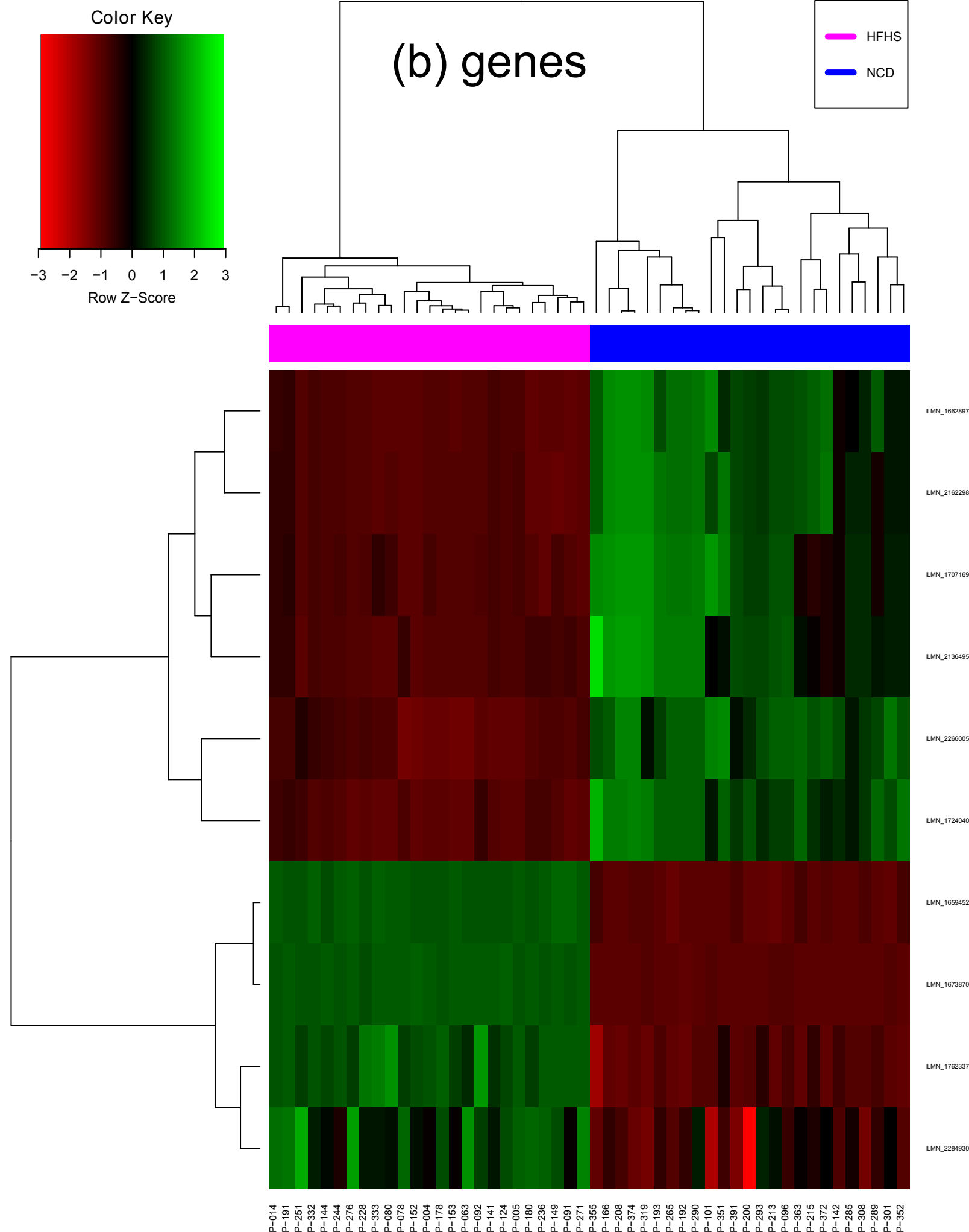

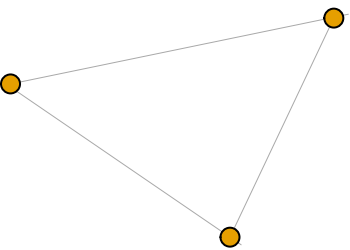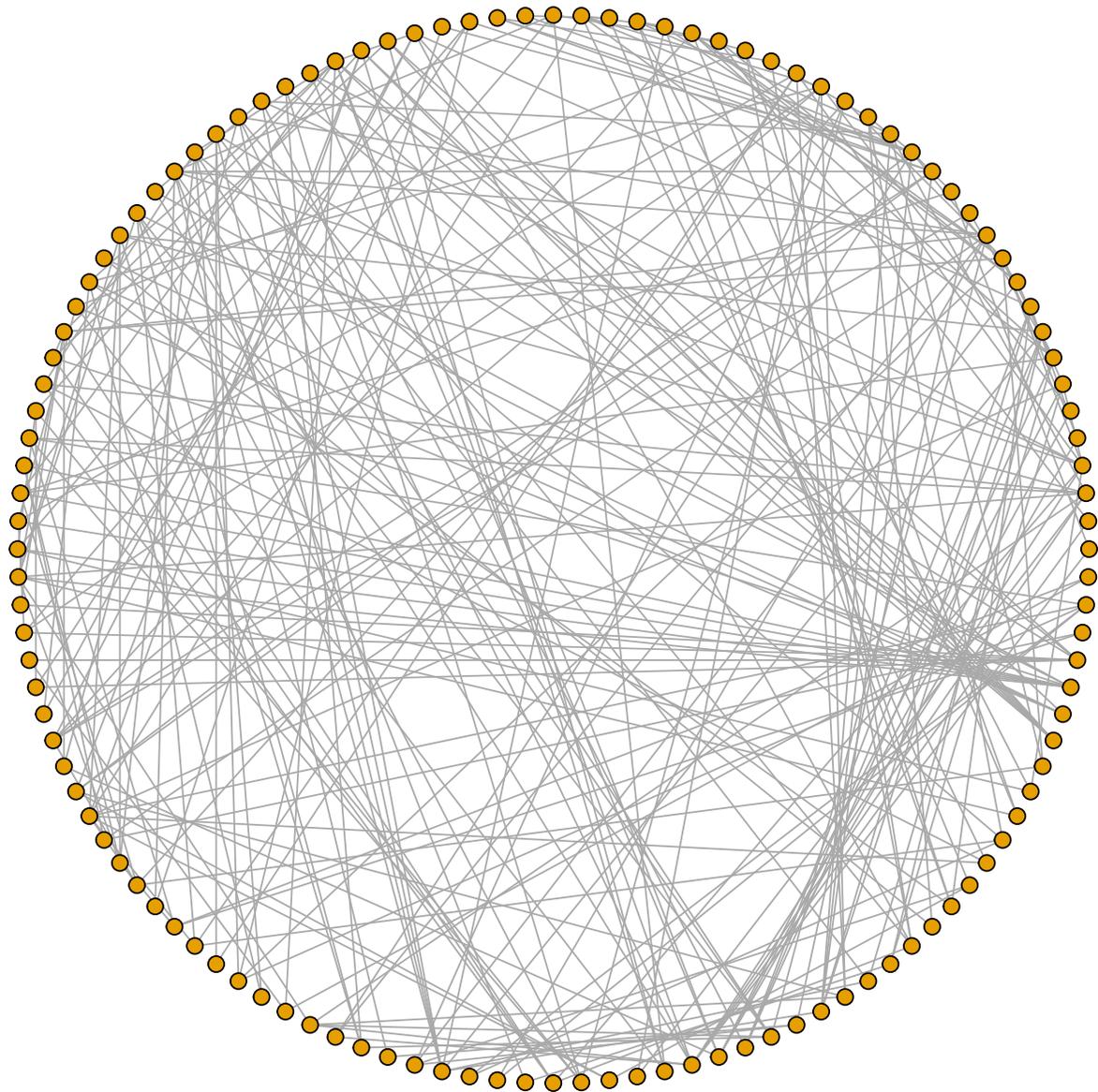

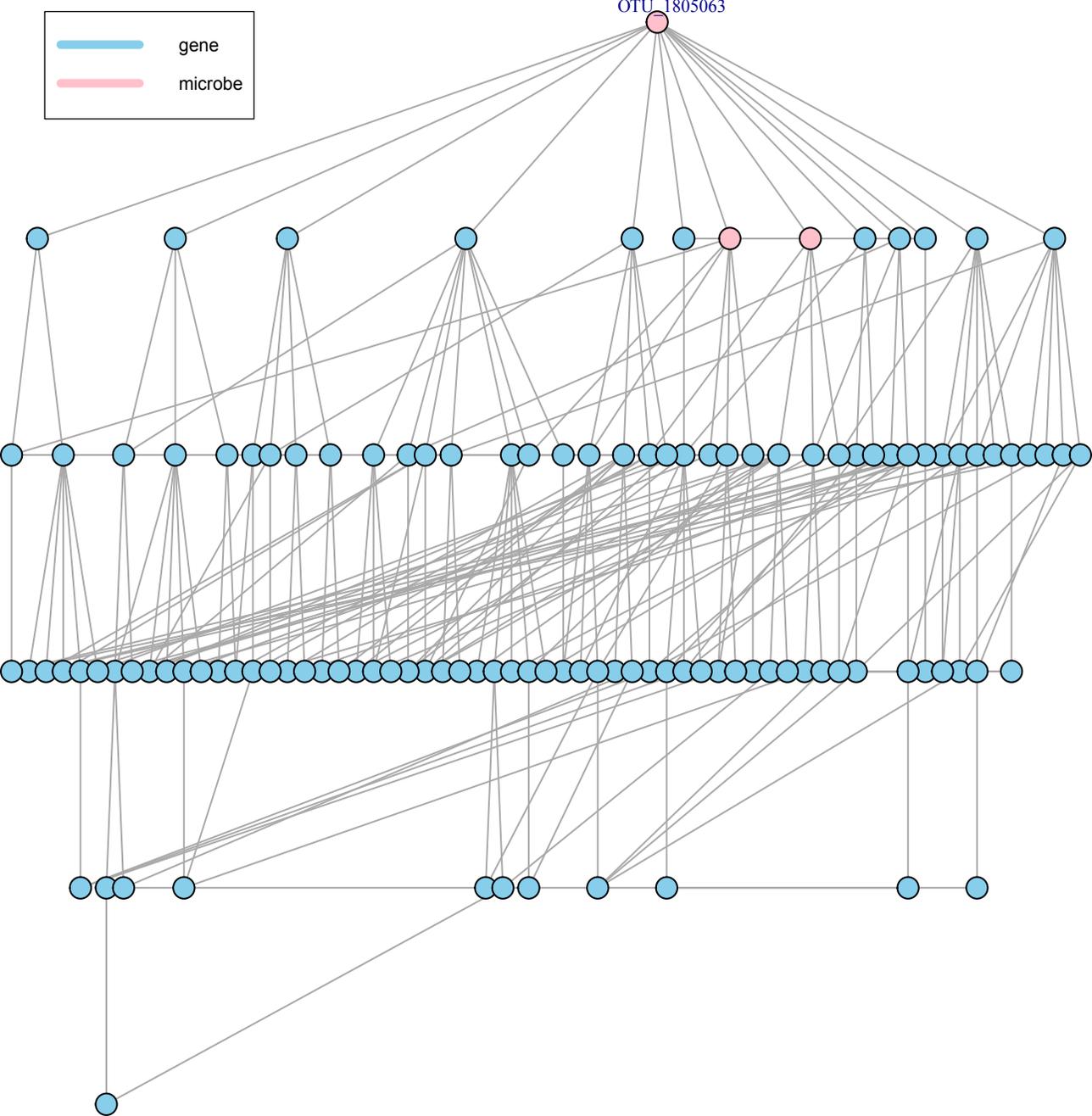

richrr / TransNetDemo

👁 Watch 1    ★ Star 0    ⑂ Fork 0

<> Code    ⊙ Issues 0    ⑂ Pull requests 0    ▣ Projects 0    ⌇ Pulse    ⊪ Graphs

Branch: master ▾    TransNetDemo / inst / demo / GeneDemo.R        Find file    Copy path

richrr Updated files after adding puc compatibility to the gene microbe network    cf07048  2 days ago

1 contributor

97 lines (76 sloc)  4.07 KB        Raw    Blame    History

```r
1   #library(TransNetDemo)
2   library(stringr)
3   library(ProNet)
4   library(igraph)
5
6
7   individualPvalueCutoff = 0.3
8   combinedPvalueCutoff = 0.05
9   combinedFDRCutoff = 0.2
10
11  # groups to compare
12  groupA = "HFHS"
13  groupB = "NCD"
14  sampleIdColName = "SampleID"
15  factorColName = "Factor"
16  geneSymbolColName ="IdSymbol"
17  # fold change column based on mean or median
18  foldchVar = "FoldChange_HFHS_NCD"   # or "FoldChange_Median_HFHS_NCD"
19
20  # map files
21  mapf1 = system.file("extdata", "mapping_file.rand.1.tsv", package = "TransNetDemo")
22  mapf2 = system.file("extdata", "mapping_file.rand.2.tsv", package = "TransNetDemo")
23
24  # gene files
25  genef1 = system.file("extdata", "gene_file_1.tsv", package = "TransNetDemo")
26  genef2 = system.file("extdata", "gene_file_2.tsv", package = "TransNetDemo")
27
28  # genes
29  Comp_genes = Compare_groups(mapf1, genef1, sampleIdColName, factorColName , groupA, groupB, geneSymbolColName)
30
31  # Select differentially abundant elements:
32  ######### (i) using a significance threshold #########
33  Sign_genes = Comp_genes[which(Comp_genes$FDR < 0.05),]
34  # this data has been saved as "Sign_genes_precomputed_1"
35  ######### ######### ######### #########
36  # OR
37  ######### (ii) in case you have multiple datasets, we recommend meta-analysis ##########
38  ## we have already run these steps and stored the data under the variable "Sign_genes_metaanalysis_precomputed" ##
39  Comp_genes1 = Comp_genes
40  Comp_genes2 = Compare_groups(mapf2, genef2, sampleIdColName, factorColName , groupA, groupB, geneSymbolColName)
41  numbDatasets=2  # number of datasets
42
43  s_df = merge(Comp_genes1, Comp_genes2, by="row.names")
44  rownames(s_df) = rownames(Comp_genes1)
45  s_df = Check_consistency(s_df, foldchVar, 1, numbDatasets)
46  comb_in_df = Calc_combined(s_df)
47  Sign_genes = Apply_sign_cutoffs(comb_in_df, individualPvalueCutoff, combinedPvalueCutoff, combinedFDRCutoff )
48
49  # if this returns TRUE, you did everything correct!
50  identical(rownames(Sign_genes), rownames(Sign_genes_metaanalysis_precomputed))
51  #write.csv(Sign_genes,"Sign_genes_File.csv", quote=FALSE)
52  ######### ######### ######### #########
53
```

```r
# gene pairs
Corr_pairs = Correlation_in_group(mapf1, genef1, sampleIdColName, factorColName, groupA,  geneSymbolColName, rownames(Sign_genes))

# Select significant correlations:
######### (i) using a significance threshold #########
Sign_pairs = Corr_pairs[which(Corr_pairs$pvalue < 0.05 & Corr_pairs$FDR < 0.1),]
######### ######### ######### #########
# OR
######### (ii) in case you have multiple datasets, we recommend meta-analysis ##########
## we have already run these steps and stored the data under the variable "Sign_genepairs_metaanalysis_precomputed" ##
Corr_pairs1 = Corr_pairs
Corr_pairs2 = Correlation_in_group(mapf2, genef2, sampleIdColName, factorColName, groupA,  geneSymbolColName, rownames(Sign_genes))

s_df = merge(Corr_pairs1, Corr_pairs2, by="row.names")
rownames(s_df) = rownames(Corr_pairs1)
s_df = Check_consistency(s_df, "Coefficient", 0, numbDatasets)
comb_in_df = Calc_combined(s_df)
Sign_pairs = Apply_sign_cutoffs(comb_in_df,individualPvalueCutoff, combinedPvalueCutoff, combinedFDRCutoff )

# if this returns TRUE, you did everything correct!
identical(rownames(Sign_pairs), rownames(Sign_genepairs_metaanalysis_precomputed))
#write.csv (Sign_pairs,"Sign_genes_pairs_File.csv", quote=FALSE)
######### ######### ######### #########

genes_df = Calc_median_val(Sign_genes, foldchVar)
identical(genes_df, Gene_df_precomputed)
pairs_df = Calc_median_val(Sign_pairs, "Coefficient")
outNetwork = Puc_compatible_network(pairs_df, genes_df)

# if this returns TRUE, you did everything correct!
identical(outNetwork[,c("partner1", "partner2")], Gene_network_precomputed[,c("partner1", "partner2")])
#write.csv (outNetwork,"gene-networkFile.csv", quote=FALSE)

cluster1 = Identify_subnetworks(outNetwork)
summary(cluster1)

# if this returns TRUE, you did everything correct!
identical(get.edgelist(cluster1), get.edgelist(genes_mcode_cluster1_precomputed)) # OR #sum(get.adjacency(cluster1) != get.adjacency(genes_
#write_graph(cluster1, "genes_mcode_cluster1_edges.txt", "ncol")

plot(cluster1, vertex.label=NA, layout= layout_in_circle, vertex.size=3)
```

richrr / TransNetDemo                                                        Watch 1    ★ Star 0    Fork 0

<> Code   Issues 0   Pull requests 0   Projects 0   Pulse   Graphs

Branch: master ▾    TransNetDemo / inst / demo / MicrobeDemo.R           Find file   Copy path

richrr Updated files after adding puc compatibility to the gene microbe network        cf07048  2 days ago

1 contributor

98 lines (76 sloc)   4.19 KB                                                  Raw   Blame   History

```r
#library(TransNetDemo)
library(stringr)
library(ProNet)
library(igraph)

individualPvalueCutoff = 0.3
combinedPvalueCutoff = 0.05
combinedFDRCutoff = 0.2

# groups to compare
groupA = "HFHS"
groupB = "NCD"
sampleIdColName = "SampleID"
factorColName = "Factor"
microbeSymbolColName ="IdSymbol"
# fold change column based on mean or median
foldchVar = "FoldChange_HFHS_NCD"   # or "FoldChange_Median_HFHS_NCD"

# map files
mapf1 = system.file("extdata", "mapping_file.rand.1.tsv", package = "TransNetDemo")
mapf2 = system.file("extdata", "mapping_file.rand.2.tsv", package = "TransNetDemo")

# microbe files
microbef1 = system.file("extdata", "microbe_file_1.tsv", package = "TransNetDemo")
microbef2 = system.file("extdata", "microbe_file_2.tsv", package = "TransNetDemo")

# microbes
Comp_microbes = Compare_groups(mapf1, microbef1, sampleIdColName, factorColName , groupA, groupB, microbeSymbolColName)

# Select differentially abundant elements:
######### (i) using a significance threshold #########
Sign_microbes = Comp_microbes[which(Comp_microbes$FDR < 0.05),]
# this data has been saved as "Sign_microbes_precomputed_1"
######### ######### ######### #########
# OR
######### (ii) in case you have multiple datasets, we recommend meta-analysis ##########
## we have already run these steps and stored the data under the variable "Sign_microbes_metaanalysis_precomputed" ##
Comp_microbes1 = Comp_microbes
Comp_microbes2 = Compare_groups(mapf2, microbef2, sampleIdColName, factorColName , groupA, groupB, microbeSymbolColName)
numbDatasets=2  # number of datasets

s_df = merge(Comp_microbes1, Comp_microbes2, by="row.names")
rownames(s_df) = rownames(Comp_microbes1)
s_df = Check_consistency(s_df, foldchVar, 1, numbDatasets)
comb_in_df = Calc_combined(s_df)
Sign_microbes = Apply_sign_cutoffs(comb_in_df, individualPvalueCutoff, combinedPvalueCutoff, combinedFDRCutoff )

# if this returns TRUE, you did everything correct!
identical(rownames(Sign_microbes), rownames(Sign_microbes_metaanalysis_precomputed))
#write.csv(Sign_microbes,"Sign_microbes_File.csv", quote=FALSE)
######### ######### ######### #########
```

```r
54
55   # microbe pairs
56   Corr_pairs = Correlation_in_group(mapf1, microbef1, sampleIdColName, factorColName, groupA,  microbeSymbolColName, rownames(Sign_microbes))
57
58   # Select significant correlations:
59   #########  (i) using a significance threshold #########
60   Sign_pairs = Corr_pairs[which(Corr_pairs$pvalue < 0.05 & Corr_pairs$FDR < 0.1),]
61   ######### ######### ######### #########
62   # OR
63   ######### (ii) in case you have multiple datasets, we recommend meta-analysis ###########
64   ## we have already run these steps and stored the data under the variable "Sign_microbepairs_metaanalysis_precomputed" ##
65   Corr_pairs1 = Corr_pairs
66   Corr_pairs2 = Correlation_in_group(mapf2, microbef2, sampleIdColName, factorColName, groupA,  microbeSymbolColName, rownames(Sign_microbes)
67
68   s_df = merge(Corr_pairs1, Corr_pairs2, by="row.names")
69   rownames(s_df) = rownames(Corr_pairs1)
70   s_df = Check_consistency(s_df, "Coefficient", 0, numbDatasets)
71   comb_in_df = Calc_combined(s_df)
72   Sign_pairs = Apply_sign_cutoffs(comb_in_df,individualPvalueCutoff, combinedPvalueCutoff, combinedFDRCutoff )
73
74   # if this returns TRUE, you did everything correct!
75   identical(rownames(Sign_pairs), rownames(Sign_microbepairs_metaanalysis_precomputed))
76   #write.csv (Sign_pairs,"Sign_microbes_pairs_File.csv", quote=FALSE)
77   ######### ######### ######### #########
78
79
80   microbes_df = Calc_median_val(Sign_microbes, foldchVar)
81   identical(microbes_df, Microbe_df_precomputed)
82   pairs_df = Calc_median_val(Sign_pairs, "Coefficient")
83   outNetwork = Puc_compatible_network(pairs_df, microbes_df)
84
85   # if this returns TRUE, you did everything correct!
86   identical(outNetwork[,c("partner1", "partner2")], Microbe_network_precomputed[,c("partner1", "partner2")])
87   #write.csv (outNetwork,"microbe-networkFile.csv", quote=FALSE)
88
89   cluster1 = Identify_subnetworks(outNetwork)
90   summary(cluster1)
91
92   # if this returns TRUE, you did everything correct!
93   identical(get.edgelist(cluster1), get.edgelist(microbes_mcode_cluster1_precomputed)) # OR #sum(get.adjacency(cluster1) != get.adjacency(mic
94   #write_graph(cluster1, "microbes_mcode_cluster1_edges.txt", "ncol")
95
96   plot(cluster1, vertex.label=NA, vertex.size=3)
97
```

🗎 richrr / **TransNetDemo**                                   ⊙ Watch  1    ★ Star  0    ⑂ Fork  0

<> Code    ⓘ Issues 0    ⑃ Pull requests 0    ▥ Projects 0    ⌁ Pulse    ‖ Graphs

Branch: master ▾ | **TransNetDemo** / inst / demo / **GeneMicrobeDemo.R**        Find file   Copy path

🯅 **richrr** added code to identify the top gene or microbe and label the importan…    97f87b8  2 days ago

**1** contributor

156 lines (118 sloc)  |  5.74 KB                                      Raw  Blame  History  🖥  ✎  🗑

```r
 1   #library(TransNetDemo)
 2   library(stringr)
 3   library(ProNet)
 4   library(igraph)
 5   library(ggplot2)
 6
 7
 8   individualPvalueCutoff = 0.3
 9   combinedPvalueCutoff = 0.05
10   combinedFDRCutoff = 0.2
11
12   # groups to compare
13   groupA = "HFHS"
14   groupB = "NCD"
15   sampleIdColName = "SampleID"
16   factorColName = "Factor"
17   gene_microbeSymbolColName ="IdSymbol"
18   # fold change column based on mean or median
19   foldchVar = "FoldChange_HFHS_NCD"   # or "FoldChange_Median_HFHS_NCD"
20
21   # map files
22   mapf1 = system.file("extdata", "mapping_file.rand.1.tsv", package = "TransNetDemo")
23   mapf2 = system.file("extdata", "mapping_file.rand.2.tsv", package = "TransNetDemo")
24
25   # gene_microbe files
26   gene_microbef1 = system.file("extdata", "gene_microbe_file_1.tsv", package = "TransNetDemo")
27   gene_microbef2 = system.file("extdata", "gene_microbe_file_2.tsv", package = "TransNetDemo")
28
29   pairs = expand.grid(V(genes_mcode_cluster1_precomputed)$name,V(microbes_mcode_cluster1_precomputed)$name)
30
31   # gene_microbe pairs
32   Corr_pairs = Correlation_in_group(mapf1, gene_microbef1, sampleIdColName, factorColName, groupA,  gene_microbeSymbolColName, NA, pairs)
33
34   # Select significant correlations:
35   ######### (i) using a significance threshold #########
36   Sign_pairs = Corr_pairs[which(Corr_pairs$pvalue < 0.05 & Corr_pairs$FDR < 0.1),]
37   ######### ######### ######### #########
38   # OR
39   ######### (ii) in case you have multiple datasets, we recommend meta-analysis ##########
40   ## we have already run these steps and stored the data under the variable "Sign_gene_microbe_pairs_metaanalysis_precomputed" ##
41   Corr_pairs1 = Corr_pairs
42   Corr_pairs2 = Correlation_in_group(mapf2, gene_microbef2, sampleIdColName, factorColName, groupA,  gene_microbeSymbolColName, NA, pairs)
43   numbDatasets=2   # number of datasets
44
45   s_df = merge(Corr_pairs1, Corr_pairs2, by="row.names")
46   rownames(s_df) = rownames(Corr_pairs1)
47   s_df = Check_consistency(s_df, "Coefficient", 0, numbDatasets)
48   comb_in_df = Calc_combined(s_df)
49   Sign_pairs = Apply_sign_cutoffs(comb_in_df,individualPvalueCutoff, combinedPvalueCutoff, combinedFDRCutoff )
50
51   # if this returns TRUE, you did everything correct!
52   identical(Sign_pairs, Sign_gene_microbe_pairs_metaanalysis_precomputed)
53   #write.csv (Sign_pairs,"Sign_gene_microbes_pairs_File.csv", quote=FALSE
```

```r
54    ######### ######### ######### #########
55
56    genes_df = rbind(Gene_df_precomputed, Microbe_df_precomputed)
57    pairs_df = Calc_median_val(Sign_pairs, "Coefficient")
58    outNetwork = Puc_compatible_network(pairs_df, genes_df)
59    #write.csv (outNetwork,"gene_microbe-networkFile.csv", quote=FALSE)
60
61
62    g = graph_from_data_frame(outNetwork,directed = F, vertices = NULL)
63    # if this returns TRUE, you did everything correct!
64    identical(get.edgelist(g), get.edgelist(Gene_Microbe_network_precomputed))
65    #write_graph(g, "gene_microbe_edges.txt", "ncol")
66
67
68    # create TK (bipartite) network using:
69    # (i) genes_mcode_cluster1_precomputed
70    # (ii) microbes_mcode_cluster1_precomputed
71    # (iii) Gene_Microbe_network_precomputed
72
73
74    net1 = get.edgelist(genes_mcode_cluster1_precomputed)
75    head(net1)
76
77    net2 = get.edgelist(microbes_mcode_cluster1_precomputed)
78    head(net2)
79
80    net3 = get.edgelist(Gene_Microbe_network_precomputed)
81    head(net3)
82
83    TK_Network = graph_from_data_frame(rbind(net1, net2, net3), directed = F)
84    print(TK_Network, e=TRUE, v=TRUE)
85
86    # if this returns TRUE, you did everything correct!
87    identical(get.edgelist(TK_Network), get.edgelist(TK_Network_precomputed))
88    #write_graph(TK_Network, "Trans_Kingdom_NetworkFile.txt", "ncol")
89    #write_graph(TK_Network, "Trans_Kingdom_NetworkFile_indices.txt")
90
91
92    # write the mapping of nodes name to indices and group
93    nodes = data.frame()
94    for (vertex in V(TK_Network)) {
95      Name = V(TK_Network)$name[vertex]
96      Id = vertex
97      Group = ""
98      if(Name %in% as.vector(net1)){
99        Group = "gene"
100     } else {
101       Group = "microbe"
102     }
103     nodes = rbind(nodes, cbind(Name, Id, Group))
104   }
105   colnames(nodes)
106   #write.table(nodes, "Trans_Kingdom_NetworkFile_nodes.txt", quote=F, row.names = F, sep=' ', col.names = T)
107
108   # calc bipartite betweenness centrality
109   # to find important genes
110   #FromNodes = as.numeric(nodes[nodes[,3]=="gene",2])
111   #ToNodes = as.numeric(nodes[nodes[,3]=="microbe",2])
112
113   # to find important microbes
114   FromNodes = as.numeric(nodes[nodes[,3]=="microbe",2])
115   ToNodes = as.numeric(nodes[nodes[,3]=="gene",2])
116
117   allPairs = expand.grid(FromNodes,ToNodes)
118   myNetwork = TK_Network
119   sumAllFractionsForAllNodes = Calc_bipartite_betweeness_centrality(allPairs, FromNodes, myNetwork)
120
121   head(sumAllFractionsForAllNodes)
122
123   # calculate node(s) with max. betweeness centrality
124   forPlot = colSums(sumAllFractionsForAllNodes)
125   topThree = sort(forPlot, decreasing = T)[1:3]
```

```r
126    TopNode = as.integer(names(topThree[1]))
127
128    TopNodeName = nodes[TopNode, "Name"]
129
130    TK_Network = set_vertex_attr(TK_Network, "type", index = nodes$Id, as.factor(nodes$Group))
131    # no labels
132    #plot(TK_Network, vertex.label = NA, layout=layout_as_tree, vertex.color=c( "pink", "skyblue")[1+(V(TK_Network)$type==1)], vertex.size=4)
133
134    # label only the important node
135    plot(TK_Network, vertex.label = ifelse(V(TK_Network)$name==TopNodeName, V(TK_Network)$name, NA),
136         layout=layout_as_tree, vertex.color=c( "pink", "skyblue")[1+(V(TK_Network)$type==1)], vertex.size=4, vertex.label.dist=0.15, vertex.la
137
138    # label only the microbe nodes
139    #plot(TK_Network, vertex.label = ifelse(nodes[which(nodes$Name==V(TK_Network)$name), "Group"]=='microbe', V(TK_Network)$name, NA),
140    #        layout=layout_as_tree, vertex.color=c( "pink", "skyblue")[1+(V(TK_Network)$type==1)], vertex.size=4, vertex.label.dist=0.15, vertex.
141
142    legend("topright",
143           legend = unique(nodes$Group),
144           col = c("skyblue" , "pink"),
145           lty= 1,
146           lwd = 5,
147           cex=.7
148    )
149
150
151
152    # plot in two rows
153    #ltypes = c(TRUE, FALSE)[1+(V(TK_Network)$type==1)]
154    #plot(TK_Network, vertex.label = NA, vertex.size=3, layout=layout_as_bipartite(TK_Network,ltypes, hgap = 500), vertex.color=c( "pink", "sky
155
```

 richrr / **TransNetDemo**                                                    Watch 1   ★ Star 0   Fork 0

Code | Issues 0 | Pull requests 0 | Projects 0 | Pulse | Graphs

Branch: master ▾ | **TransNetDemo** / inst / demo / **Heatmaps.R**        Find file   Copy path

richrr finalized code with legends                                            643b5e8  19 days ago

1 contributor

66 lines (49 sloc) | 1.92 KB                                     Raw   Blame   History

```r
1   #library(TransNetDemo)
2   library(gplots)
3
4
5   ########################### extract inputs - study 1 #############################
6
7   # groups to compare
8   groupA = "HFHS"
9   groupB = "NCD"
10  sampleIdColName = "SampleID"
11  factorColName = "Factor"
12  geneSymbolColName ="IdSymbol"
13
14  # map files
15  mapf1 = system.file("extdata", "mapping_file.rand.1.tsv", package = "TransNetDemo")
16
17  # read the mapping file
18  map1 = read.delim(mapf1, header=T)
19  rownames(map1) = map1[,sampleIdColName]
20
21  # select samples from groups
22  hfhs_samples1 = as.vector(map1[which(map1[,factorColName] == groupA),sampleIdColName])
23  ncd_samples1 = as.vector(map1[which(map1[,factorColName] == groupB),sampleIdColName])
24
25  # colors as per the group
26  samplecolors <- c(rep("magenta",25) , rep("blue",25))
27
28  # gene files
29  genef1 = system.file("extdata", "gene_file_1.tsv", package = "TransNetDemo")
30
31  # read the data file
32  genes1 = read.delim(genef1, header=T, check.names = F, row.names=1)
33
34  # make the heat map
35  dataset1=genes1[rownames(Sign_genes_precomputed_1), c(hfhs_samples1, ncd_samples1)]
36  dim(dataset1)
37  heatmap.2(as.matrix(dataset1), col=redgreen(75), ColSideColors=samplecolors, scale="row", key=TRUE, symkey=FALSE, density.info="none", trac
38
39  legend("topright",
40         legend = unique(map1$Factor),
41         col = c("magenta" , "blue"),
42         lty= 1,
43         lwd = 5,
44         cex=.7
45  )
46
47  # microbe files
48  microbef1 = system.file("extdata", "microbe_file_1.tsv", package = "TransNetDemo")
49
50  # read the data file
51  microbes1 = read.delim(microbef1, header=T, check.names = F, row.names=1)
52
53  # make the heat map
```

```
54    dataset1=microbes1[rownames(Sign_microbes_precomputed_1), c(hfhs_samples1, ncd_samples1)]
55    dim(dataset1)
56    heatmap.2(as.matrix(dataset1), col=redgreen(75), ColSideColors=samplecolors, scale="row", key=TRUE, symkey=FALSE, density.info="none", trac
57
58    legend("topright",
59           legend = unique(map1$Factor),
60           col = c("magenta" , "blue"),
61           lty= 1,
62           lwd = 5,
63           cex=.7
64    )
65
```

richrr / **TransNetDemo**

 Watch  1      ★ Star  0      ⑂ Fork  0

<> Code   ⓘ Issues 0   ⑃ Pull requests 0   ▯ Projects 0   ⚡ Pulse   ⅠⅡ Graphs

Branch: master ▾   **TransNetDemo** / **R** / **Apply_sign_cutoffs.R**   Find file   Copy path

 **richrr** Updated package                             f30fa7f  26 days ago

**1** contributor

24 lines (22 sloc) | 1.23 KB                 Raw   Blame   History

```r
 1  #' Select significant elements
 2  #'
 3  #' This function applies the following significance cutoffs
 4  #' individual p-value < 0.3; combined p-value < 0.05; fdr < 0.1
 5  #'
 6  #' @param df, individualPvalueCutoff, combinedPvalueCutoff, combinedFDRCutoff
 7  #' @return A matrix with statistics for signifcant elements
 8  #' @export
 9  Apply_sign_cutoffs = function(df, individualPvalueCutoff = 0.3, combinedPvalueCutoff = 0.05, combinedFDRCutoff = 0.1){
10    # find significant in individual pvalue: all of the pvalues for all of the datasets for each element must be smaller than threshold
11    # find pvalue data
12    # if it is a single dataset it becomes a vector so adding the drop=FALSE
13    pvalueData = df[,grep("pvalue",colnames(df)), drop=FALSE]
14    print(head(pvalueData))
15    pvalueData = as.matrix(pvalueData)
16    pvalueData = apply(pvalueData,2,function(x){as.numeric(as.vector(x))})
17
18    # calculate the largest pvalue among all datasets for each gene, this largest pvalue must be smaller than threshold
19    passIndevidualPvalue = apply(pvalueData,1,max)<individualPvalueCutoff
20    Sign_elements = df[passIndevidualPvalue,]
21    Sign_elements <- Sign_elements[Sign_elements$"combinedPvalue" < combinedPvalueCutoff  &  Sign_elements$"combinedFDR" < combinedFDRCutoff,
22    return(Sign_elements)
23  }
```

richrr / **TransNetDemo**    Watch 1    ★ Star 0    Fork 0

Code | Issues 0 | Pull requests 0 | Projects 0 | Pulse | Graphs

Branch: master ▾ | **TransNetDemo** / R / **Calc_bipartite_betweeness_centrality.R**    Find file    Copy path

richrr Updated package    f30fa7f 26 days ago

1 contributor

32 lines (26 sloc) | 1020 Bytes    Raw    Blame    History

```r
 1  #' Calc_bipartite_betweeness_centrality
 2  #'
 3  #' This function calculates bipartite_betweeness_centrality for each given element pair
 4  #'
 5  #' @param allPairs, FromNodes, myNetwork
 6  #' @return A data frame with statistics for each element
 7  #' @export
 8  Calc_bipartite_betweeness_centrality = function(allPairs, FromNodes, myNetwork){
 9
10    sumAllFractionsForAllNodes = Get_template_matrix(FromNodes, allPairs)
11    sumAllFractionsForAllNodes = as.data.frame(sumAllFractionsForAllNodes)
12    #print(sumAllFractionsForAllNodes)
13    counts = apply(as.matrix(allPairs),1,Get_shortest_paths,myNetwork)
14    #print(counts)
15
16
17    z = lapply(counts, function(x){
18      if(nrow(x)>0){
19        #print(x)
20        columnsForUpdate = as.character(intersect(colnames(x),FromNodes))
21        #print(columnsForUpdate)
22        rowsForUpdate = rownames(x)
23        #print(rowsForUpdate)
24        #print(x[1, columnsForUpdate])
25        sumAllFractionsForAllNodes[rowsForUpdate,columnsForUpdate] <<- x[1,columnsForUpdate]
26      }
27    })
28
29    return(sumAllFractionsForAllNodes)
30  }
31
```

richrr / **TransNetDemo**                      Watch  1    ★ Star  0    Fork  0

<> Code  ⓘ Issues 0  ⑂ Pull requests 0  ▦ Projects 0  ⟋⟍ Pulse  ▮▮ Graphs

Branch: master ▾ | **TransNetDemo** / R / **Calc_combined.R**                Find file   Copy path

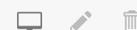 **richrr** Updated package                                                     f30fa7f 26 days ago

**1** contributor

29 lines (26 sloc) | 1.02 KB                                    Raw   Blame   History   🖥  ✎  🗑

```r
 1  #' Calc combined p value across expts.
 2  #'
 3  #' This function code for calc combined p value across datasets.
 4  #'
 5  #' @param data frame
 6  #' @return A matrix with combined pvalue and fdr
 7  #' @export
 8  Calc_combined = function(s_df){
 9    pvalueData = s_df[,grep("pvalue",colnames(s_df)), drop=FALSE]
10    head(pvalueData)
11
12    total_numb_input_files = ncol(pvalueData)
13    interestedPvalueData = as.matrix(pvalueData)
14    interestedPvalueData = apply(interestedPvalueData,2,function(x){as.numeric(as.vector(x))})
15    combinedPvalue = apply(interestedPvalueData,1
16                           ,function(pvalues){
17                             pvalues = pvalues[!is.na(pvalues)]
18                             statistics = -2*log(prod(pvalues))
19                             degreeOfFreedom = 2*length(pvalues)
20                             combined = 1-pchisq(statistics,degreeOfFreedom)
21                           }
22    )
23    #calculate FDR for combined pvalue
24    combinedFDR = p.adjust(combinedPvalue,method="fdr")
25    comb_in_df = cbind(s_df, combinedPvalue, combinedFDR)
26
27    return(comb_in_df)
28  }
```

Watch  1    Star 0    Fork 0

Code | Issues 0 | Pull requests 0 | Projects 0 | Pulse | Graphs

Branch: master ▾  TransNetDemo / R / **Calc_cor.R**    Find file    Copy path

**richrr** Updated package    f30fa7f 26 days ago

**1** contributor

19 lines (15 sloc) | 425 Bytes    Raw  Blame  History

```r
#' Calculate correlation between vectors
#'
#' This function calculate correlation between vectors
#'
#' @param pair, df, samples
#' @return A matrix with statistics for each element pair
#' @export
Calc_cor = function(pair, df, samples){
  idxs = as.vector(samples)
  
  c1 = as.numeric(df[pair[1], idxs])
  c2 = as.numeric(df[pair[2], idxs])
  
  p = cor.test(c1,c2)
  outLine = as.matrix(c(p$estimate,p$p.value))
  outLine
}
```

📖 richrr / **TransNetDemo**    👁 Watch  1    ★ Star  0    ⑂ Fork  0

<> Code  ⓘ Issues 0  ⑂ Pull requests 0  📁 Projects 0  ⚡ Pulse  📊 Graphs

Branch: master ▾ | **TransNetDemo** / R / **Calc_median_val.R**    Find file   Copy path

**richrr** Updated package    f30fa7f 26 days ago

**1** contributor

20 lines (19 sloc) | 723 Bytes    Raw  Blame  History  🖥 ✏ 🗑

```r
 1  #' Calculate median coefficient or fold change
 2  #'
 3  #' This function calculate median coefficient or fold change for each element
 4  #'
 5  #' @param df, pattern
 6  #' @return A data frame with statistics for each element
 7  #' @export
 8  Calc_median_val = function(df, pattern){
 9    Colnames = colnames(df)[grep(pattern,colnames(df))]
10    interestedData = df[,Colnames,drop=F]
11    interestedData = as.matrix(interestedData)
12    interestedData = apply(interestedData,2,function(x){as.numeric(as.vector(x))})
13    combined = apply(interestedData,1, function(x){round(median(x, na.rm = TRUE), 3)})
14    oldColnames = colnames(df)
15    df = cbind(df,combined)
16    colnames(df) = c(oldColnames, paste("combined", pattern, sep=''))
17    print(head(df))
18    return(df)
19  }
```

 richrr / **TransNetDemo**   Watch 1   ★ Star 0   Fork 0

Code · Issues 0 · Pull requests 0 · Projects 0 · Pulse · Graphs

Branch: master ▾   TransNetDemo / R / **Check_consistency.R**   Find file   Copy path

**richrr** Updated package   f30fa7f 26 days ago

**1** contributor

23 lines (19 sloc)  855 Bytes   Raw  Blame  History

```
1  #' Check consistent direction
2  #'
3  #' This function selects elements showing consistent direction across datasets
4  #'
5  #' @param df, patternToSearch, threshold
6  #' @return A matrix with elements showing consistent direction across datasets
7  #' @export
8  Check_consistency = function(s_df, pattern, Threshold, numbDatasets=2){
9    # select the data of interest
10   patternData = s_df[,grep(pattern,colnames(s_df)), drop=FALSE]
11   head(patternData)
12
13   rows_passing_consistency = c()
14   res_pos = apply(patternData, 1, function(x) sum(x > Threshold))
15   rows_passing_consistency = c(rows_passing_consistency, names(res_pos[res_pos==numbDatasets]))
16
17   res_neg = apply(patternData, 1, function(x) sum(x < Threshold))
18   rows_passing_consistency = c(rows_passing_consistency, names(res_neg[res_neg==numbDatasets]))
19
20   s_df = s_df[rows_passing_consistency, -1]
21   return(s_df)
22 }
```

📖 richrr / **TransNetDemo**    ⦿ Watch 1   ★ Star 0   ⑂ Fork 0

<> Code    ⓘ Issues 0    ⇡⇣ Pull requests 0    ▣ Projects 0    ⚡ Pulse    📊 Graphs

Branch: master ▾ | **TransNetDemo** / R / **Compare_groups.R**    Find file   Copy path

➕ **richrr** Updated package    f30fa7f 26 days ago

**1** contributor

45 lines (34 sloc) | 1.69 KB    Raw   Blame   History   🖥   ✏   🗑

```r
#' Compare groups
#'
#' This function compares the values in two groups for each element
#'
#' @param mapFile, geneFile, sampleIdColName, factorColName, groupA, groupB, geneSymbolColName
#' @return A data frame with statistics for each element
#' @export
Compare_groups = function(mapFile, geneFile, sampleIdColName, factorColName, groupA, groupB, geneSymbolColName){

  ########################### extract inputs ###########################
  # read the mapping file
  map = read.delim(mapFile, header=T)
  rownames(map) = map[,sampleIdColName]
  head(map)

  # select samples from groups
  hfhs_samples = map[which(map[,factorColName] == groupA),]
  head(hfhs_samples)
  ncd_samples = map[which(map[,factorColName] == groupB),]
  head(ncd_samples)

  # read the data file
  genes = read.delim(geneFile, header=T, check.names = F)
  rownames(genes) = genes[,geneSymbolColName]
  head(genes)

  ########################### calc diff abundance ###########################
  out = sapply(rownames(genes), Diff_abundance, genes, rownames(hfhs_samples), rownames(ncd_samples))

  ########################### format output ###########################
  Comp_genes = t(out)
  colnames(Comp_genes) = c(paste("Mean", groupA, sep="_"), paste("Mean", groupB, sep="_"), paste("FoldChange", groupA, groupB, sep="_"),
                  paste("Median", groupA, sep="_"), paste("Median", groupB, sep="_"), paste("FoldChange","Median", groupA, groupB, sep="_

  ########################### calculate FDR ###########################
  FDR = p.adjust(Comp_genes[,"pvalue"],method="fdr")
  oldColnames = colnames(Comp_genes)
  Comp_genes = as.data.frame(cbind(Comp_genes, FDR))

  return(Comp_genes)
}
```

richrr / **TransNetDemo**

◉ Watch 1 | ★ Star 0 | ⑂ Fork 0

<> Code | ⊙ Issues 0 | ⑂ Pull requests 0 | ⊞ Projects 0 | ⚡ Pulse | ▮ Graphs

Branch: master ▾ | **TransNetDemo** / R / **Correlation_in_group.R** | Find file | Copy path

richrr Updated package | f30fa7f 26 days ago

**1** contributor

54 lines (42 sloc) | 1.86 KB | Raw | Blame | History

```r
#' Correlation_in_group
#'
#' This function calculates correlation in a group for each element pair
#'
#' @param mapFile, geneFile, sampleIdColName, factorColName, groupA, geneSymbolColName, selected_genes
#' @return A data frame with statistics for each element
#' @export
Correlation_in_group = function(mapFile, geneFile, sampleIdColName, factorColName, groupA,  geneSymbolColName, selected_genes, usepairs=NA)

  ########################### extract inputs ###############################
  # read the mapping file
  map = read.delim(mapFile, header=T)
  rownames(map) = map[,sampleIdColName]
  head(map)

  # select samples from group
  hfhs_samples = map[which(map[,factorColName] == groupA),]
  head(hfhs_samples)

  # read the data file
  genes = read.delim(geneFile, header=T, check.names = F)
  rownames(genes) = genes[,geneSymbolColName]
  head(genes)

  # create gene pairs if not already provided
  pairs = ''
  if(is.na(usepairs)) {
    pairs = t(combn(selected_genes,2))[,2:1]
  }else{
    pairs = usepairs
  }
  head(pairs)

  ########################## calculate correlation for each pair ###############################
  Corr_pairs = apply(pairs, 1, Calc_cor, genes, rownames(hfhs_samples))

  ########################## format output ###############################
  # add the pair as the column name
  colnames(Corr_pairs) = paste(as.vector(pairs[,1]),as.vector(pairs[,2]),sep="<==>")
  # the pairs become row names; and the corr coeff and pvalue are the column names
  Corr_pairs = t(Corr_pairs)
  # append method used to column labels
  colnames(Corr_pairs) = c(paste(groupA, "Coefficient",sep="_"), "pvalue")

  ########################## calculate FDR  ###############################
  FDR = p.adjust(Corr_pairs[,colnames(Corr_pairs)[grep("pvalue",colnames(Corr_pairs))]],method="fdr")
  Corr_pairs = as.data.frame(cbind(Corr_pairs, FDR))

  return(Corr_pairs)
}
```

**Personal** **Open source** **Business** **Explore** Pricing Blog Support [This repository] [Search] Sign in Sign up

📖 richrr / **TransNetDemo**  👁 Watch 1  ★ Star 0  ⑂ Fork 0

<> Code | Issues 0 | Pull requests 0 | Projects 0 | Pulse | Graphs

Branch: master ▾ | TransNetDemo / R / Diff_abundance.R    Find file  Copy path

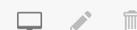 richrr Updated package    f30fa7f 26 days ago

**1** contributor

29 lines (22 sloc) | 790 Bytes    Raw  Blame  History

```r
 1  #' Compare abundance between groups for an element
 2  #'
 3  #' This function compares the values in two groups for an element
 4  #'
 5  #' @param element, df, samplesA, samplesB
 6  #' @return A matrix with statistics for each element
 7  #' @export
 8  Diff_abundance = function(element, df, samplesA, samplesB){
 9    samplesA = as.vector(samplesA)
10    samplesB = as.vector(samplesB)
11
12    Vec1 = as.numeric(df[element, samplesA])
13    Vec2 = as.numeric(df[element, samplesB])
14
15    meanVec1 = mean(Vec1)
16    meanVec2 = mean(Vec2)
17    fold_change_mean = meanVec1/meanVec2
18
19    medianVec1 = median(Vec1)
20    medianVec2 = median(Vec2)
21    fold_change_median = medianVec1/medianVec2
22
23    p = t.test(Vec1, Vec2)
24
25    Result = as.matrix(c(meanVec1, meanVec2, fold_change_mean, medianVec1 , medianVec2 , fold_change_median, p$p.value))
26
27    Result
28  }
```

Personal  Open source  Business  Explore    Pricing  Blog  Support   This repository  Sign in  Sign up

richrr / **TransNetDemo**    Watch 1   ★ Star 0   Fork 0

`<> Code`   Issues 0   Pull requests 0   Projects 0   Pulse   Graphs

Branch: master ▾ | **TransNetDemo** / **R** / **Get_shortest_paths.R**    Find file  Copy path

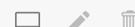 **richrr** Updated package    f30fa7f  26 days ago

**1** contributor

34 lines (30 sloc) | 1.58 KB    Raw  Blame  History

```r
#' Get counts for all shortest paths between given nodes in a bipartite networks
#'
#' This function calculates shortest paths between given nodes in a bipartite networks
#'
#' @param pair, myNetwork, sumAllFractionsForAllNodes
#' @return Updates the row in pair,sumAllFractionsForAllNodes
#' @export
Get_shortest_paths = function(pair,myNetwork){  # for each row in "sumAllFractionsForAllNodes", calculate the values for the columns and up
  #print(pair)
  allShortestPaths = get.all.shortest.paths(myNetwork, from= pair[1], to= pair[2], mode = "all", weights=NULL) # calculate the shortest pat
  #print(allShortestPaths)

  count = data.frame()
  if (length(allShortestPaths$res)==0){  # if there is no shortest path, do not update "sumAllFractionsForAllNodes"
    #fractions = cbind(FromNodes,rep(0,times=length(FromNodes)))
    #print("no paths")
  }else{
    allShortestPaths = do.call(rbind,allShortestPaths$res) # a matrix with each row containing a shortest path between the two nodes
    #print(allShortestPaths)

    nodesInPath = as.vector(allShortestPaths[,c(-1,-ncol(allShortestPaths))]) # get rid of the two nodes that are under study, the nodes in
    #print(nodesInPath)
    count = table(as.factor(nodesInPath))/nrow(allShortestPaths) # for each node in the shortest paths, calculate how many times(normalized
    #print(count)

    v = names(count)
    count = data.frame(rbind((count)))
    colnames(count) = v
    rownames(count) = paste(pair, collapse='_')
    #print(count)
  }
  count
}
```

richrr / **TransNetDemo**   👁 Watch  1   ★ Star  0   ⑂ Fork  0

<> Code    Issues 0    Pull requests 0    Projects 0    Pulse    Graphs

Branch: master ▾    TransNetDemo / R / **Get_template_matrix.R**    Find file   Copy path

**richrr** Updated package    f30fa7f 26 days ago

**1** contributor

22 lines (21 sloc)  1.05 KB    Raw  Blame  History

```
 1  #' Get_template_matrix
 2  #'
 3  #' This function creates a matrix , with each entry recording for each node(in a column),
 4  #' if it appears(not 1 or 0, but a normalized number, since there can be more
 5  #' than one shortest path between a pair of nodes) in the shortest path between a
 6  #' pair of nodes(in a row), the rows contain all pairs of nodes between two groups.
 7  #' Thus the sum of each column is the betweeness centrality for the node(in that column)
 8  #'
 9  #' @param FromNodes, allPairs
10  #' @return A matrix with rownames (source-target) and column names as nodes from source nodes set.
11  #' @export
12  Get_template_matrix = function(FromNodes, allPairs){
13    sumAllFractionsForAllNodes = matrix(0,ncol=length(FromNodes)
14                                        ,nrow=nrow(allPairs)
15                                        ,dimnames= list(
16                                          paste(as.vector(t(allPairs[,1])),as.vector(t(allPairs[,2])),sep="_")
17                                          , as.character(FromNodes))
18                                        )
19    )
20    return(sumAllFractionsForAllNodes)
21  }
```

richrr / **TransNetDemo**                                        ⊙ Watch  1    ★ Star  0    ⑂ Fork  0

<> Code    ⊙ Issues 0    ⑂ Pull requests 0    ⊟ Projects 0    ⌁ Pulse    ⊞ Graphs

Branch: master ▾   TransNetDemo / R / **Identify_subnetworks.R**                  Find file   Copy path

⊕ **richrr** Updated package                                                 f30fa7f 26 days ago

**1** contributor

37 lines (30 sloc)  1.16 KB                              Raw  Blame  History  ▭ ✎ 🗑

```r
 1  #' Identify top subnetwork
 2  #'
 3  #' This function returns the induced subgraph of the nodes in the top cluster using mcode
 4  #'
 5  #'
 6  #' @param network in which to identify subnetworks
 7  #' @return A data frame (network) that is Puc compatiable
 8  #' @export
 9  Identify_subnetworks = function(outNetwork){
10
11    # identify subnetworks
12    dfNetwork = outNetwork[,c("partner1", "partner2", "combinedCoefficient")]
13    print(head(dfNetwork))
14
15    #?graph_from_data_frame
16    # http://kateto.net/networks-r-igraph
17    g = graph_from_data_frame(dfNetwork,directed = F, vertices = NULL)
18    print(g, e=TRUE, v=TRUE)
19    # see the edges and vertices
20    #E(g) ; V(g) ; edge_attr(g) ; vertex_attr(g)
21
22    # plots the networks
23    #cluster(g,method="MCODE",layout="fruchterman.reingold")
24    #cluster(g,method="FN",layout="fruchterman.reingold")
25
26    #https://cran.r-project.org/web/packages/ProNet/vignettes/Tutorial.pdf
27    # identify clusters
28    result <- mcode(g,vwp=0.5,haircut=TRUE,fluff=FALSE,fdt=0.8,loops=FALSE)
29    summary(result$COMPLEX)
30
31    # plot the top cluster
32    cluster1<-induced.subgraph(g,result$COMPLEX[[1]])
33    #visualization(cluster1,node.size=4,node.label=V(cluster1)$name,node.label.color="blue")
34
35    return(cluster1)
36  }
```

Branch: master ▾   **TransNetDemo** / **R** / **Puc_compatiable_network.R**   Find file   Copy path

richrr Updated package   f30fa7f 26 days ago

**1** contributor

83 lines (66 sloc) | 3.69 KB   Raw   Blame   History

```r
1   #' Calculate network that satisfies the PUC criteria
2   #'
3   #' This function keeps edges that satisfy expected correlation and fold change relationships
4   #'
5   #'
6   #' @param pairs_df, genes_df
7   #' @return A data frame (network) that is Puc compatiable
8   #' @export
9   Puc_compatible_network = function(pairs_df, genes_df){
10    #---------------------------------------------------------------
11    # calculate PUC
12    #---------------------------------------------------------------
13
14    #change out format to partner1 partner2 for PUC
15    row_names_pairs_df = rownames(pairs_df)
16    head(row_names_pairs_df)
17    pair = stringr::str_split( row_names_pairs_df ,"<==>")
18    pairs = t(as.data.frame(pair))
19
20    colnames(pairs) = c("partner1","partner2")
21    pairs_df = apply(pairs_df, 2, function(x) as.numeric(as.character(x))) # convert the chars to numeric
22    rownames(pairs) = row_names_pairs_df  # remove this if you do not want row names
23    outForPUC = cbind(pairs,pairs_df)
24    head(outForPUC)
25
26
27    # select pvalue columns
28    grep_cols_c = grep("pvalue", colnames(outForPUC), value=TRUE, fixed=TRUE)
29    # select "combinedPvalue", "combinedFDR", "combinedCoefficient"
30    grep_cols_c = append(grep_cols_c, grep("combined" , colnames(outForPUC), value=TRUE, fixed=TRUE) )
31    # select the partner columns
32    g_grep_cols = c("partner1","partner2", grep_cols_c)
33
34    outForPUC = outForPUC[,g_grep_cols]
35    head(outForPUC)
36
37
38    # attach the foldChange information for each partner
39    FoldChangeCol = grep(paste("combined" , foldchVar, sep=''), colnames(genes_df), value=TRUE, fixed=TRUE)
40    FoldMetab1_InPair = genes_df[as.vector(outForPUC[,"partner1"]), FoldChangeCol, drop=F]
41    colnames(FoldMetab1_InPair) = c("partner1_FoldChange")
42    FoldMetab2_InPair = genes_df[as.vector(outForPUC[,"partner2"]), FoldChangeCol, drop=F]
43    colnames(FoldMetab2_InPair) = c("partner2_FoldChange")
44    outForPUC = cbind(outForPUC,FoldMetab1_InPair,FoldMetab2_InPair)
45    head(outForPUC)
46
47    # calculate correlation Direction For combined correlation coefficient of interest
48    # at this point we only have the consistent pairs left, so the value of combined corr coeff is ok to use
49    interestedCoefficientColnames = grep("Coefficient",colnames(outForPUC), value=TRUE, fixed=TRUE)
50    print(interestedCoefficientColnames)
51    interestedCorrelationData = outForPUC[,interestedCoefficientColnames, drop=FALSE]
52    interestedCorrelationData = apply(interestedCorrelationData,2,function(x){as.numeric(as.vector(x))})
53    correlationDirection = interestedCorrelationData/abs(interestedCorrelationData)
```

```r
54
55
56      # calculate fold change direction for each partner
57      FoldChangeColnames = colnames(outForPUC)[grep("FoldChange",colnames(outForPUC))] # since this is using the combined fold change calculate
58
59      FoldChangeData = outForPUC[,FoldChangeColnames]
60      FoldChangeDirection = (FoldChangeData-1)/abs(FoldChangeData-1)
61      names(FoldChangeDirection) = c()
62      colnames(FoldChangeDirection) = paste(colnames(FoldChangeData),"Direction",sep="_")
63
64      # calculate if fold change direction are the same for the two partners
65      IfFoldChangeDirectionMatch = apply(FoldChangeDirection,1,prod)
66      names(IfFoldChangeDirectionMatch) = c()
67      colnames(correlationDirection) = c()
68
69      # use "correlationDirection" and "IfFoldChangeDirectionMatch" to calc PUC,
70      # i.e. if these two are the same PUC=1 (good)
71      PUC = IfFoldChangeDirectionMatch * correlationDirection
72      outForPUC = cbind(outForPUC,correlationDirection,FoldChangeDirection,IfFoldChangeDirectionMatch,PUC)
73      head(outForPUC)
74
75      #write.csv(outForPUC, "PUC-output.csv" ,row.names=FALSE)
76
77      # find PUC expected
78      out = outForPUC[outForPUC[,"PUC"]==1 ,]
79      outNetwork = out[!is.na(out$"PUC"),] # remove the rows with 'NA' in PUC columns
80      return(outNetwork)
81
82  }
```